\documentclass[twocolumn]{aastex63}\tighten
\usepackage{lineno}
\usepackage{amsmath}
\received{2022 September 23}
\revised{2023 January 24}
\accepted{2023 January 25}

\submitjournal{ApJ}

\shorttitle{NIR polarimetry of outflow source S235\,e2s3}
\shortauthors{Devaraj et al.}

\def\Msun{\hbox{$M_{\odot}$}}
\def\Lsun{\hbox{$L_{\odot}$}}
\newcommand{\htwo}{H$_2$}
\newcommand{\wm}{${\rm W\,m}^{-2}$}

\begin{document}

\title{\large Near-Infrared Polarimetry and H$_2$ emission toward Massive Young Stars:\\ Discovery of a Bipolar Outflow associated to S235\,e2s3 }

\correspondingauthor{Devaraj Rangaswamy}
\email{dev2future@yahoo.com}

\author[0000-0001-9217-3168]{R. Devaraj}
\affiliation{Dublin Institute for Advanced Studies, 31 Fitzwilliam Place, Dublin D02XF86, Ireland}

\author[0000-0001-8876-6614]{A. Caratti o Garatti}
\affiliation{Istituto Nazionale di Astrofisica, 16 Salita Moiariello, Naples 80131, Italy}
\affiliation{Dublin Institute for Advanced Studies, 31 Fitzwilliam Place, Dublin D02XF86, Ireland}

\author[0000-0001-6725-0483]{L. K. Dewangan}
\affiliation{Physical Research Laboratory, Navrangpura, Ahmedabad, Gujarat 380009, India}

\author[0000-0003-4040-4934]{R. Fedriani}
\affiliation{Department of Space, Earth and Environment, Chalmers University of Technology, 412 96 Gothenburg, Sweden}

\author[0000-0002-2110-1068]{T. P. Ray}
\affiliation{Dublin Institute for Advanced Studies, 31 Fitzwilliam Place, Dublin D02XF86, Ireland}

\author[0000-0002-3922-6168]{A. Luna}
\affiliation{Instituto Nacional de Astrof\'isica, \'Optica y Electr\'onica, Tonantzintla, Puebla 72840, M\'exico}

%\author{et al.}

\begin{abstract} %currently 278 words

We present a near-infrared $H$ band polarimetric study toward the S235\,e2s3 protostar, obtained using the POLICAN instrument on the 2.1m OAGH telescope. The images reveal a bipolar outflow with a total length of about 0.5\,pc. The outflow nebulosity presents a high degree of linear polarization ($\sim$80\%) and reveals a centrosymmetric pattern with the polarization position angles. The polarization characteristics suggest their origin to be single scattering associated with dust in the outflow. Using multiwavelength archival data, we performed spectral energy distribution (SED) fitting based on radiative transfer models of turbulent core accretion theory. The best-fit SED model indicated that the protostar has a mass of $6.8\pm1.2$\,\Msun, with a disk accretion rate of $3.6\pm1.2\times10^{-4}\,\mathrm{M_\odot\,yr^{-1}}$ and a total bolometric luminosity of $9.63\pm2.1\times10^{3}$\,\Lsun. Narrowband H$_2$ ($2.12\,\mu$m) observations show shocked emission along the bipolar lobes tracing the jet's interaction with the surrounding medium. The estimated \htwo\ luminosity of the outflow is $2.3_{-1.3}^{+3.5}\,\Lsun$, which matched the known power-law correlation with the source bolometric luminosity, similar to other high-mass outflows. The orientation of the bipolar outflow was found to be parallel to the local magnetic field direction. The overall results assert the fact that the S235\,e2s3 source is a massive young star driving a highly collimated bipolar outflow through disk accretion.

\end{abstract}


\section{Introduction} \label{sec1}

\begin{figure*}[t!]
\epsscale{1.2}
\plotone{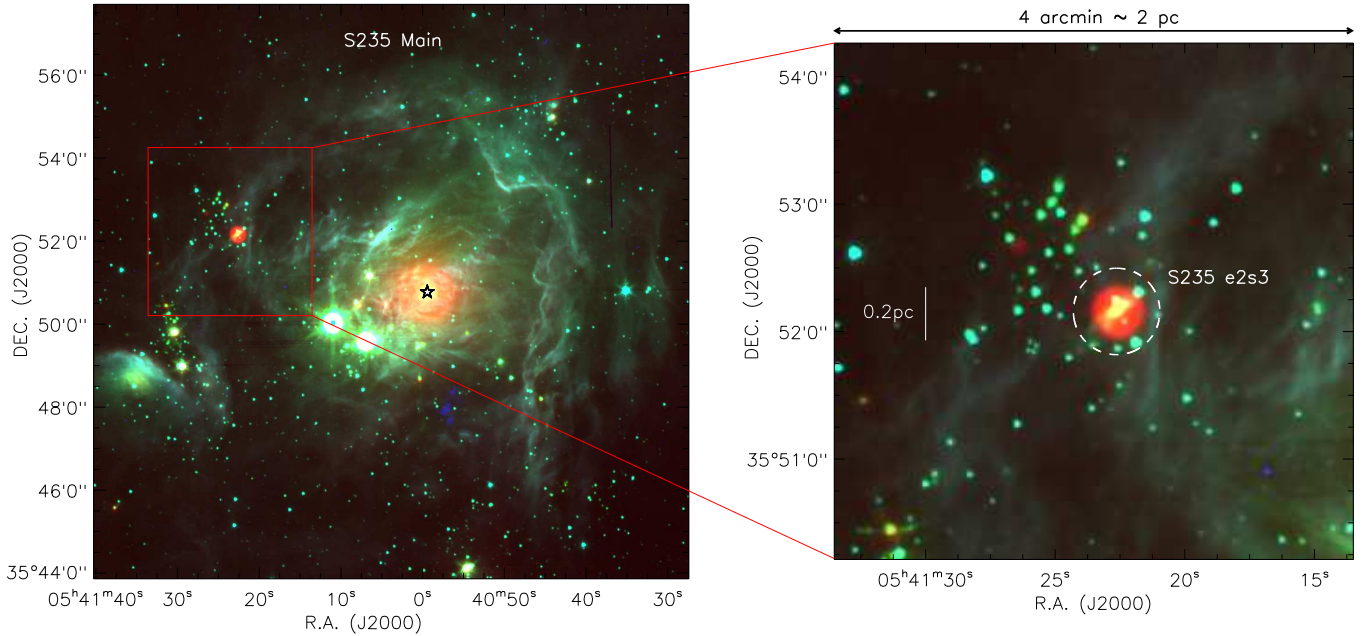}
\caption{MIR three-color image of the S235 complex, constructed using \textit{Spitzer} MIPS and IRAC bands: $24\,\mu$m (red), $4.5\,\mu$m (green) and $3.6\,\mu$m (blue) \citep{fazio04}. The left panel shows the larger S235~Main region spanning $15\arcmin \times 14\arcmin$, corresponding to a physical scale of $7.2\,\mathrm{pc}\times 6.7\,\mathrm{pc}$ at a distance of $1.65\,\rm{kpc}$. The central ionizing O9.5V star for S235~Main is marked by the star symbol. The right panel shows a zoomed-in view of the region centered on the S235~e2s3\,source, spanning a $4\arcmin \times 4\arcmin$ FOV or a physical scale of about $2\,\mathrm{pc}\times 2\,\mathrm{pc}$. The dashed white circle encompasses the S235\,e2s3 source.}
\label{fig1}
\end{figure*}

Molecular outflows are one of the most prominent signposts of the star formation process. They are intimately connected to accretion and ejection phenomena of the embedded protostar \citep{pudritz83,ray07, bally16}. The exact mechanism driving these outflows is unclear, but according to models they can be due to a disk wind \citep{blan82,shu94} or interactions between the disk field and protostellar magnetosphere \citep{shang07,shu07} or by collimated flows from toroidal magnetic fields \citep{ban06,ban07}. A unifying factor in all these mechanisms is their dependence on magnetic field interaction with the flows \citep{frank14}. Recent observations suggest that the formation of intermediate- to high-mass stars also proceeds via disk-mediated accretion as with their low-mass counterparts, powering highly collimated outflows \citep{davis04, gredel06, fedriani18}. The outflows hold insights about the associated protostars and can reveal the nature of their origins through their morphologies, kinematics, and physical properties. One way to the study the interaction of these flows is through molecular H$_{2}$ emission which is a particularly good shock tracer as it is the primary coolant in the near-infrared (NIR) and its emission can extend over large spatial scales \citep{caratti06, caratti08}. Hence, observations of outflows with different techniques are crucial in constraining the evolution of intermediate and high-mass stars.

NIR imaging polarimetry offers a unique opportunity to study the morphology and characteristics of outflows. The nebulosity associated to the outflows presents high degrees of linear polarization that can only be attributed to light scattering from dust particles \citep{werner83,bastien88}. These observations can reveal information on the spatial distribution of the scattering material and the inclination of the outflow \citep{tamura91, jones04, meakin05}. The composition and size distribution of the dust grains responsible for scattering can also be investigated using polarimetry \citep[e.g.][]{pendleton90, kim94}. Generally, the young stars driving the most powerful outflows are invisible at optical and NIR wavelengths as they are embedded in dust. Polarimetric observations allow us to identify and locate the position of these illuminating stars through their outflow polarization signatures \citep{tamura97,weintraub00}.

The object of study in this work is `S235\,e2s3', which is located in the eastern subregion of the S235~complex (see Figure~\ref{fig1}). The S235 is a well studied star-forming complex harboring an evolved H\,{\sc ii} region: `S235~Main' \citep{kirsanova08, dewangan11, dewangan16}. The H\,{\sc ii} region is ionized by a central massive star of O9.5V spectral type \citep{georgelin73}. The region is estimated to be at a distance of $1.65_{-0.10}^{+0.12}\,\rm kpc$ based on the Gaia DR2 \citep{brown18} parallax of the ionizing star. The molecular gas in the surrounding environment has been traced with velocities ranging from $-22\,\mathrm{kms^{-1}}$ to $-20\,\mathrm{kms^{-1}}$ and has been shown to present a number of star forming clusters \citep{kirsanova08,kirsanova14,chavarria14}. Various studies have concluded that the clusters are a result of positive stellar feedback driven by the expansion of the S235~Main H\,{\sc ii} region \citep{dewangan16,dewangan17,dev21}.

The source S235\,e2s3 was identified by \citet{dewangan11} through a \textit{Spitzer} IRAC \citep{fazio04} mid-infrared (MIR) photometric study of the S235 complex. They inferred it to be a class 0/I young stellar object (YSO) lacking any NIR counterparts. \citet{chavarria14} used MIR color-color diagrams and identified the source as a class I YSO. The object spatially coincides less than $30\arcsec$ with the IRAS 05379+3550 source and is listed in the \citet{lumsden2013} catalog of massive young stars derived from the MSX survey \citep{egan03}. Preliminary spectral energy distribution (SED) modeling of the source by \citet{dewangan11} revealed it to be a high-mass protostellar object. Further studies of IRAC ratio maps ($4.5\,\mu$m/$3.6\,\mu$m) and narrowband H$_{2}$ images (2.12$\,\mu$m) by \citet{dewangan16} showed shocked and molecular emission surrounding the central source. The maps also revealed an extended region with elongated structures, indicating the presence of an outflow. These results suggest that the S235\,e2s3 protostar presents active accretion and ejection processes. To understand better the physical properties and the nature of the extended structures, we have carried out a new NIR imaging polarimetric study toward the region of the embedded protostar.

\begin{figure*}[th!]
\epsscale{1.22}
\plotone{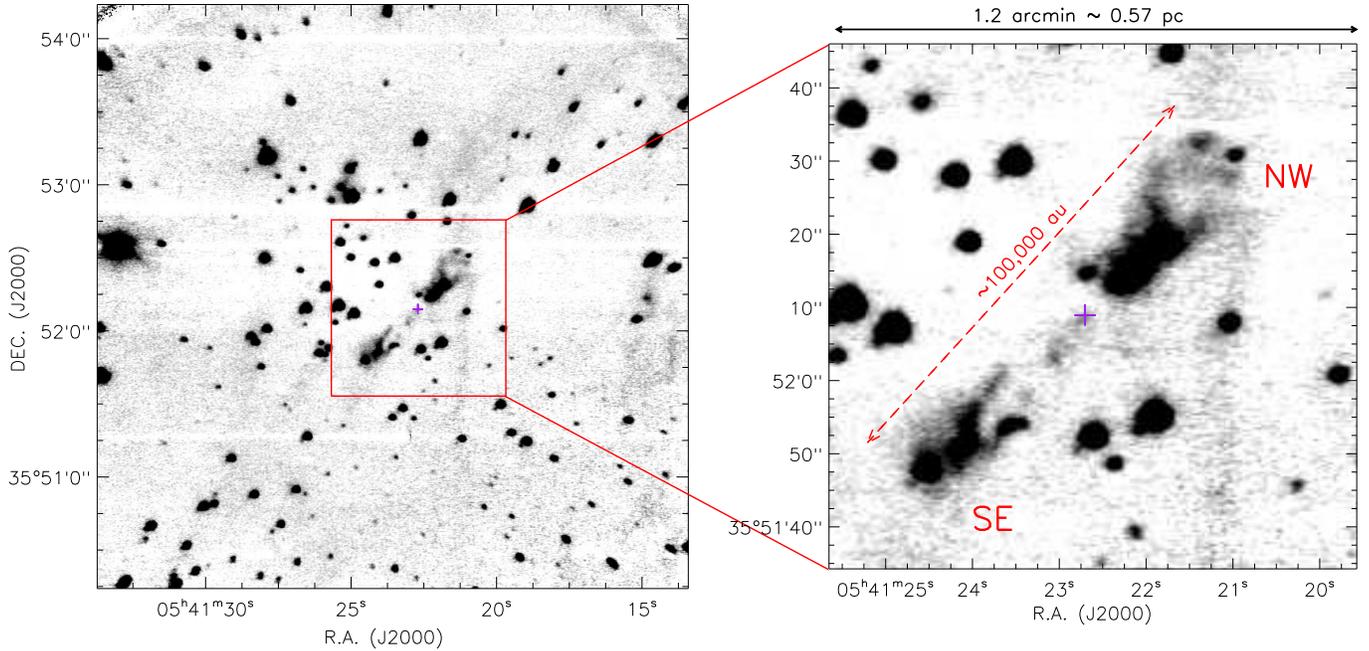}
\caption{NIR gray-scale image of the total intensity emission obtained in $H$ band from the polarimetric observations. The image was constructed by average-combining a set of 18 images resulting in a total integration time of 2 hrs. The left panel image spans a $4\arcmin \times 4\arcmin$ FOV of the POLICAN instrument. The right panel shows a zoomed-in view of the S235\,e2s3 outflow, spanning a $1.2\arcmin \times 1.2\arcmin$ field equivalent to a physical scale of about 0.57\,pc. The two bipolar lobes are marked NW and SE. The total length of the bipolar nebulosity ($\sim100,000$\,au) is shown as a red dashed line. The line also corresponds to the position angle of the outflow axis ($137^{\circ}$). In both the panels the location of the S235\,e2s3 protostar is marked by a purple cross.}
\label{fig2}
\end{figure*}

The paper is organized as follows: in Section 2, we describe details of the observations and data reduction techniques. In Section 3, we present polarization maps and show the spatial distribution of the outflow. In Section 4, we discuss the results about the morphology of the outflow, the polarization properties, magnetic fields, and the \htwo\ luminosity. Finally in Section 5, we summarize the work with concluding remarks.

\section{Observations and Data Sets} \label{sec2}

\subsection{NIR polarization observations} \label{NIRobs}
NIR linear polarimetric observations were obtained in $H$ band ($1.6\,\mu$m) using the POLICAN instrument \citep{dev18b} attached to the CANICA camera \citep{car17} on the $2.1\,\mathrm{m}$ OAGH telescope in Cananea, Sonora, Mexico. The camera has a HgCdTe detector with 1024$\times$1024 pixels and a pixel-scale of $0.32\,\mathrm{arcsec}$. The field of view (FOV) achieved with the instrument is $4\arcmin\times4\arcmin$. The observations were acquired on the 26$^{\rm th}$ of September 2017, centered on the S235\,e2s3 source with coordinates $\alpha_{2000}$ = 05$^{h}$41$^{m}$22.705$^{s}$, $\delta_{2000}$ = +35$\degr$52$\arcmin$08.93$\arcsec$ \citep{dewangan11}. The average atmospheric seeing during the observations was about $0\arcsec.9$. The angular size of the $4\arcmin$ FOV corresponds to a physical scale of about $2\,\mathrm{pc}$ at a distance of $1.65\,\mathrm{kpc}$.

Polarimetric imaging was performed through combination of a rotating half-wave plate (HWP) as the modulator and a fixed wire-grid polarizer as the analyser. In order to measure the linear polarization, the HWP is rotated to four position angles $(0^{\circ}, 22$.$5^{\circ}, 45^{\circ}, 67$.$5^{\circ})$ to complete one set of images. Each set of images were obtained at 18 sky dithered positions to estimate the sky contribution and boost the signal-to-noise ratio when the images were combined. The exposure times were set to $100\,\mathrm{s}$ per image, totaling an integration time of $2\,\mathrm{hrs}$ for the 72 images (18$\times$4~HWP).

Image reduction consisted of dark subtraction, flat-fielding, and sky removal. After basic reduction, the dithered images were aligned and combined for each HWP angle. They were then astrometry-corrected using data from 2MASS \citep{skrutskie06}. The pixel intensities resulting from the four combined HWP images were used to compute the Stokes parameters $I$, $Q$, and $U$ for the field \citep[see][]{dev18b}. 
 
The polarization degree ($P$) and position angle ($PA$) were computed from the Stokes parameters, respectively, as follows:
$P = \sqrt{Q^{2}+U^{2}}/I$,\,\, $PA = \frac{1}{2}\tan^{-1}\left(\frac{U}{Q}\right)$.
The polarization uncertainty ($\sigma_{P}$) was also computed from the corresponding Stokes errors. Since $P$ is positively biased from the quadrature combination of $Q$ and $U$, the de-biased degree of polarization ($P^{'}$; also expressed as percentage polarization) was estimated using a Ricean correction as $P^{'} = \sqrt{P^{2} - \sigma_{P}^{2}}$ \citep{wardle74}. The foreground polarization contribution from interstellar dust toward the S235 region is minimal at $\sim0.3\%$ \citep{dev21}. Hence, the removal of foreground polarization is neglected in the analysis. Full descriptions of the calibration procedures and polarimetric analyses are given in \citet{dev18a,dev18b}. The final polarization values at each pixel were stored as an image array and were binned for every 3$\times$3 pixels to create the polarization maps for the observed region.

\subsection{Archival data sets}  
  
Additional archival data sets, ranging from NIR to radio wavelengths, were used in our study. MIR images were obtained from Spitzer IRAC and MIPS data sets through guaranteed time observations (GTOs) \citep{fazio04}. IRAC images (at 3.6, 4.5, 5.8, and $8\,\mu$m) have an angular resolution of $\sim2\arcsec$ and the MIPS $24\,\mu$m images have $6\arcsec$ resolution. Narrowband H$_{2}$ ($\nu=1-0$ S(1); $\lambda/\Delta\lambda = 2.122\,\mu$m/0.032\,$\mu$m) imaging data were obtained from the extended H$_{2}$ emission survey \citep{navarete15}, conducted using WIRcam at the CFHT. The survey also provided $K$ band continuum images, which were used to get the final continuum-subtracted H$_{2}$ map. 

For the purpose of SED modeling, NIR $K$ band photometric data were taken from 2MASS \citep{skrutskie06}. MIR photometric fluxes were taken from the WISE all sky release (3.4, 4.6, 12, $22\,\mu$m) \citep{wright10}, MSX Galactic plane survey (8.3, 12.1, 14.7, 21.3$\,\mu$m) \citep{egan03} and AKARI point source catalog \citep{abrahamyan15}. Far-IR and millimeter fluxes (at 350$\mu$m and 1.1\,mm) were taken from the extended Bolocam Galactic Plane Survey, version 2.1 \citep{aguirre11, ginsburg13, merello15}. James Clark MAxwell Telescope (JCMT) Scuba observations were used for the flux at 850$\,\mu$m \citep{klein05}.

\section{Analyses and Results} \label{sec3}

\subsection{NIR total intensity and Polarization Maps}

The NIR $H$ band total intensity image from the observations is shown in Figure~\ref{fig2}. The image is centered on the S235\,e2s3 source (indicated by a purple cross) and extends to a size of $4\arcmin \times 4\arcmin$, equivalent to a physical scale of $2\,\mathrm{pc}\times 2\,\mathrm{pc}$. The most prominent result seen in the image is the two large extended structures of nebulosity appearing to trace an outflow. The nebulosity clearly shows its distinction as a collimated outflow and differs from the typical IR reflection nebulosity due to a dusty envelope around a compact H\,{\sc ii} region. A zoomed-in view of the region toward the nebulosity is shown in the right panel of Figure~\ref{fig2}. The two structures are diametrically opposite to each other and form the morphology of bipolar outflow lobes. One lobe is oriented north-west, whereas the other is pointed to the south-east. These are marked as NW and SE, respectively. The nebulosity in the NW lobe is much closer to the central source when compared to the SE lobe.
We determined a rough value for the position angle of the outflow axis by bisecting the outflow lobes around their spatial distribution. The position angle was estimated to be around $137^{\circ}$ (measured from the north-up to east-left direction). The red dashed line in Figure~\ref{fig2} is shown parallel to the outflow axis. There is no significant NIR emission in the central region between the lobes. This could be due to the fact that the central source is deeply embedded in circumstellar material and not visible in our $H$ band observations. 

\begin{figure*}[th!]
\epsscale{1.15}
\plotone{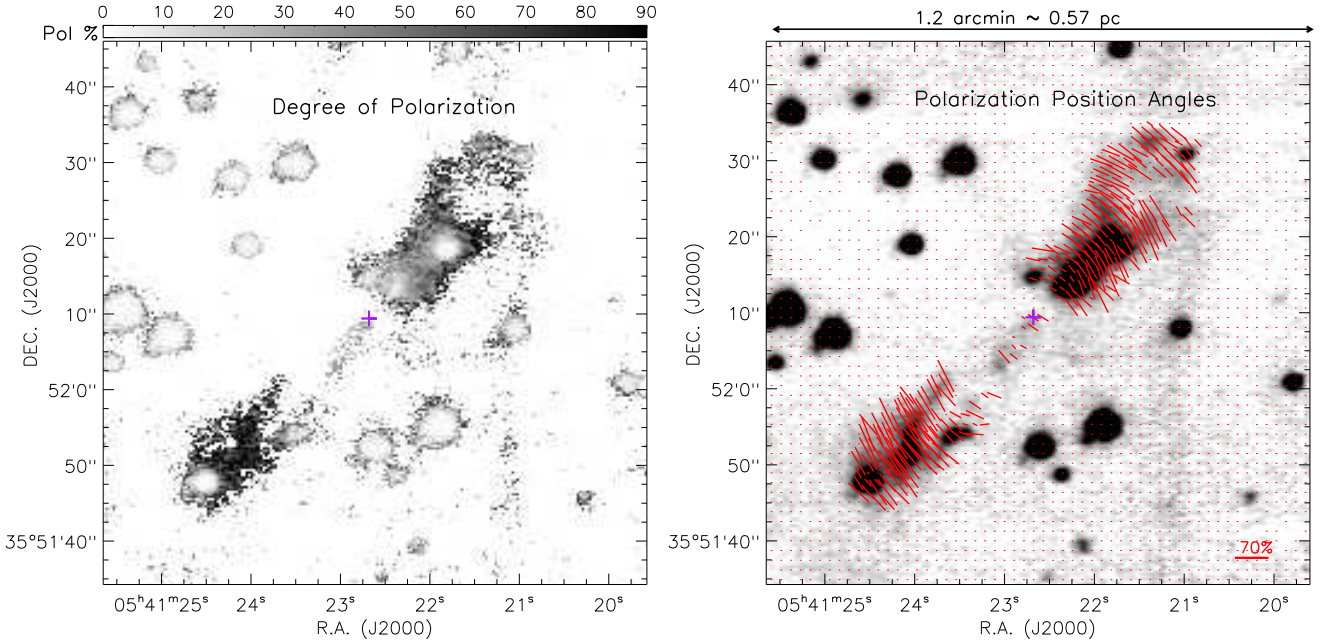}
\caption{Polarization maps of the S235\,e2s3 bipolar outflow. The left panel shows the de-biased degree of polarization. A gray-scale color bar at the top shows the percentage range of the degree of polarization. The right panel shows the polarization position angles represented by red vectors. The lengths of the vectors correspond to the degree of polarization. The polarization angles display a centrosymmetric pattern around the S235\,e2s3 source, which is marked by a purple cross in both panels.}
\label{fig3}
\end{figure*}

The polarimetric imaging results are presented in Figure~\ref{fig3} as maps of the degree of polarization and position angles. The maps have been constructed using polarization values estimated at each pixel as described in Section~\ref{NIRobs}. The left panel in Figure~\ref{fig3} shows a gray-scale image of the degree of polarization with the color bar at the top indicating the range of values. Darker regions correspond to higher polarization and lighter regions for lower polarization. The polarization levels in the region are remarkably high and reach a maximum of about 80$\%$. The bipolar nebulosity is the most prominent structure seen in the polarized flux. The regions associated to the nebulosity show high degree of polarization across both lobes, averaging about 70$\%$. A few foreground/background stars toward the outflow are revealed in the polarization map that are not visible in the total intensity image in Figure~\ref{fig2}. The stars have significantly lower polarizations of less than 10\%. There is no significant polarization emission at the location of the central source. A streamer-like elongated structure is visible from the source center toward the SE lobe with polarization values of up to 30\%. This could be from dust in or around the jet from the young star. In general, the high polarization levels in the nebulosity show excellent correlation, spatially matching the morphology of the outflow seen in Figure~\ref{fig2}.

The distribution of polarization position angles is shown as red vectors in the right panel of Figure~\ref{fig3}. Only the regions associated to the nebulosity having a polarization level threshold above 3$\sigma$ are plotted. The most notable feature in the map is the centrosymmetric pattern of the polarization vectors. The pattern is equally distributed in both bipolar lobes. This feature together with the very high degree of polarization is indicative that the nebulosity is being illuminated by a single source at its center. The light from the illuminating source is singly scattered at near right angles into our line of sight. Such scattering mechanism has been associated with regions with low optical depth \citep{gledhill01, gledhill05}. The polarization at the central source location (near the purple cross) is faint and the polarization vectors seem to appear perpendicular to the axis of the outflow. This polarization pattern could originate from the dusty disk around the young star (see Section~\ref{polprop} for more details). However, there is no direct indication of the presence of a disk, as our observations lack the resolution to resolve one. Overall, the results of the polarization maps with the centrosymmetric pattern suggest that the NW and SE lobes belong to the same region and emanate from a single source. 

\subsection{Narrowband \texorpdfstring{H$_{2}$}\, image} \label{h2sec}

\begin{figure*}[t!]
\epsscale{1.1}
\plotone{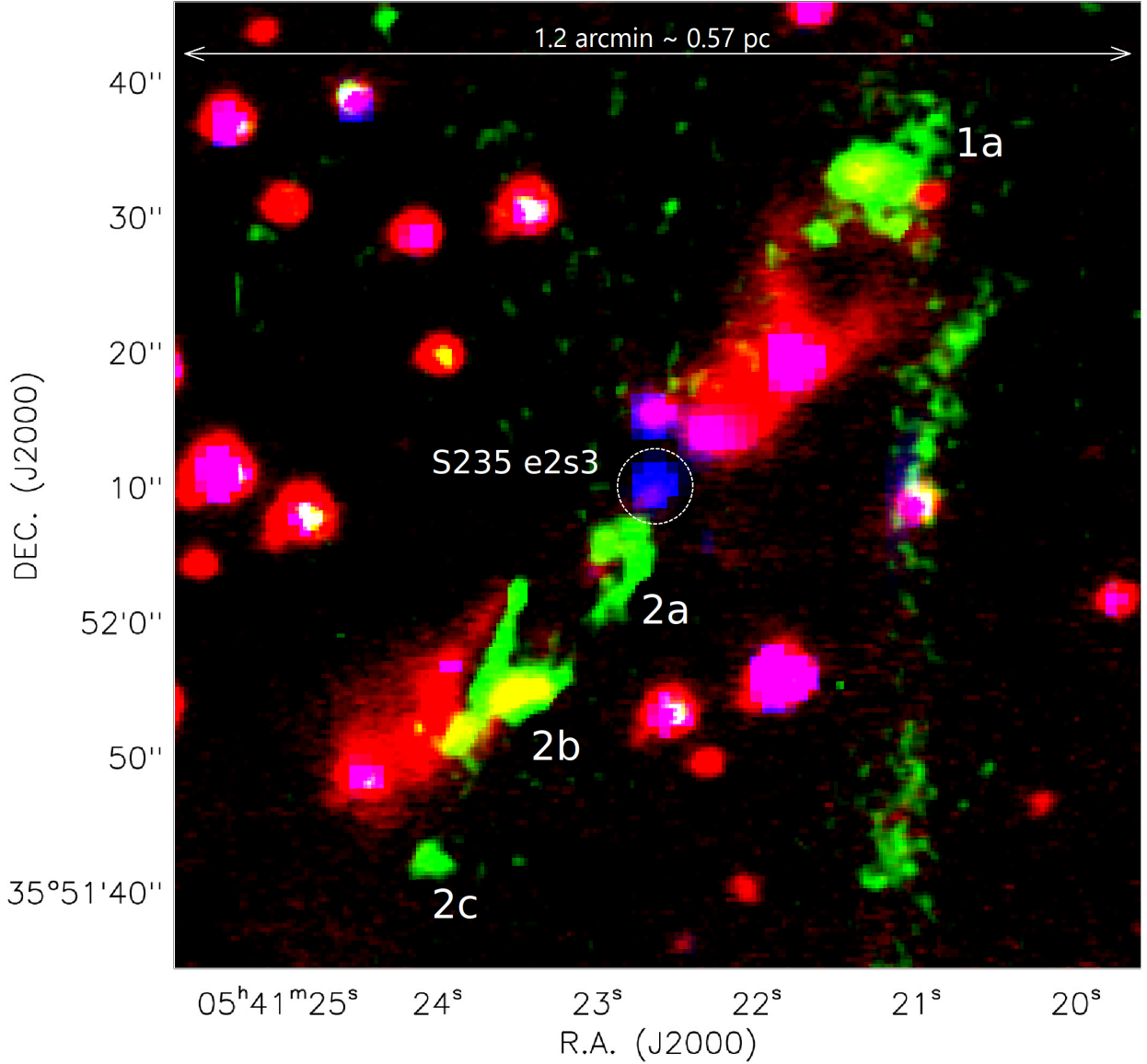}
\caption{Three-color composite map of the outflow region, constructed using $H$ band 1.6$\,\mu$m (red), H$_{2}$ emission 2.12$\,\mu$m (green), Spitzer 4.5$\,\mu$m (blue). The image spans $1.2\arcmin \times 1.2\arcmin$, corresponding to a projected physical scale of $0.57\,\mathrm{pc}\times 0.57\,\mathrm{pc}$. The H$_{2}$ emission in green with its multiple components are marked as 1a, 2a, 2b, and 2c. The S235\,e2s3 source is seen in blue from the Spitzer 4.5$\,\mu$m image and is outlined by a dashed circle.}
\label{fig4}
\end{figure*}

Observations in NIR narrowband H$_{2}$ emission trace shock-heated gas in molecular outflows and are useful to investigate the ejection processes in young stars \citep{navarete15, kim18}. In Figure~\ref{fig4} we present a three-color composite image of the outflow region using NIR continuum-subtracted H$_{2}$ emission at 2.12$\,\mu$m (green), NIR $H$ band at 1.6$\,\mu$m (red), and Spitzer 4.5$\,\mu$m (blue). The displayed H$_{2}$ emission is $3\sigma$ above the background. We identified several H$_{2}$ emission features toward the nebulosity which are composed of discrete ﬂows/shocks. These are named as 1a, 2a, 2b, and 2c in Figure~\ref{fig4}. The most remarkable result is that the H$_{2}$ emission is bipolar and very closely coincides with the axis of the outflow in the plane of the sky. The shocked emission is narrow and well-collimated within the opening angles of the outflow. The extent of H$_{2}$ emission is similar to the dusty nebulosity and reaches to $\sim19\arcsec$ or about 31,000\,au from the center. The maximum cross sectional width is $\sim6\arcsec$ or about 9,900\,au.

The NW outflow apparently consists of single shocked H$_{2}$ emission area at its tip (marked as 1a). The SE region has two distinct elongated emissions. One very close to the central source (marked as 2a) and the other located before the SE dusty lobe (marked as 2b). The 2b H$_{2}$ emission coincides exceptionally well with the boundary of the dusty nebulosity. This region particularly reveals a scenario consisting of a bow shock and the different layers associated with it. There is a small emission region located away from the southern tip of the nebulosity (marked as 2c). This emission does not coincide directly with the outflow axis, but is within $10\arcsec$ of the outflow region. The positional offset in the 2c emission could be due to several reasons, such as projection effects or precession of the jet. We do not find any further \htwo\ emission related to the outflow beyond our FOV when comparing the large-scale \htwo\ map of the region \citep{dewangan16}. We note that in Figure~\ref{fig4} toward the west of the nebulosity, there is a long vertical feature with diffuse H$_{2}$ emission that is not spatially associated to the outflow. This feature is due to UV-excited emission (fluorescence) from the S235~main ionizing star (see Figure~\ref{fig1} for details).
The overall morphology of the H$_{2}$ emission, along with the NIR dusty nebulosity and the position of the central source (blue in Figure~\ref{fig4}) exhibit a well-defined ejection process occurring with the S235\,e2s3 source. 

\subsection{SED modeling}\label{SED}

\begin{table}
\movetabledown=13mm
\footnotesize
\centering
\caption{Fluxes from NIR to radio wavelengths used for the SED modeling.}
\label{tab1}
%\hspace*{-2cm}
\begin{tabular}{llll}
\hline 
\hline
 Facility/ & Wavelength & Flux Density$^{\mathrm{a}}$ & Comments \\
 Instrument & $\lambda\,(\mu \rm{m})$ &  (Jy) & \\
\hline

2MASS K 	  		& 2.2    & 0.46 (0.08) $\times10^{-3}$  & upper limit \\
WISE: W1 	  		& 3.4    & 7.87 (0.37) $\times10^{-3}$	& upper limit \\
WISE: W2	  		& 4.6    & 34.2 (0.8) $\times10^{-3}$  & upper limit \\
Spitzer/Ch3  		& 5.8    & 51.3 (0.1) $\times10^{-3}$  & upper limit \\
Spitzer/Ch4 		& 8.0    & 105 (0.04) $\times10^{-3}$ 	& \\
MSX/A		  		& 8.3    & 0.20 (0.01) & \\
WISE: W3	  		& 11.6   & 0.62 (0.01)	& \\
MSX/C		  		& 12.13  & 0.96 (0.11) 	& \\
MSX/D		  		& 14.6   & 1.16 (0.1) 	& \\
AKARI/IRC-L	  		& 18	 & 4.50 (0.17)	& \\
MSX/E		  		& 21.3   & 5.29 (0.34) 	& \\
WISE: W4		  	& 22.1   & 11.18 (0.2)	& \\
AKARI/FIS-N60 		& 65	 & 178 (8.13)	&  \\
AKARI/FIS-L      	& 140	 & 230 (105) 	& \\
CSO/Bolocam	 		& 350	 & 19.95 (0.38) & \\
JCMT/SCUBA   		& 850    & 1.38 (0.52)  & \\
CSO/Bolocam	 		& 1100	 & 1.04 (0.12)	& \\

\hline          
\end{tabular}
\tablecomments{$^{\mathrm{a}}$The values in the parenthesis are the flux errors.}

\end{table}

The SED modeling of the S235\,e2s3 source was carried out to derive its physical parameters and properties of the mass accreting disk/envelope. \citet{dewangan11} previously carried out preliminary SED analysis for the S235\,e2s3 source with limited data points using \citet{robitaille07} SED fitting. However, the \citet{robitaille07} SED models were developed mainly for low-mass protostars without considering massive star formation properties such as high accretion rates and high-mass surface densities. \citet{debuizer17} studied a sample of massive protostars from the SOFIA survey and showed that \citet{robitaille07} models typically result in high-mass infall rates but about 100 times smaller disk accretion rates, leading to smaller bolometric luminosities and a larger protostellar masses.

Hence, here we use SED fitting tool \textit{`SedFitter'} that is part of the python package \verb+sedcreator+\footnote{https://github.com/fedriani/sedcreator} implemented by \citet{fedriani23}. The SED fitting code is developed using radiative transfer grid models of intermediate and high-mass stars from \citet{zhang18}, based on the turbulent core accretion theory of massive star formation \citep{mckee03}. The initial conditions in the massive star formation model are pressurized, dense, massive cores embedded in high-mass surface density clump environments. These massive stars are assumed to form from preassembled massive prestellar cores supported by internal pressure from a combination of magnetic fields and turbulence. In the model grid there are three main parameters that set the physical properties: the initial mass of the core ($M_c$), the mean mass surface density of the clump environment ($\Sigma_\mathrm{cl}$), and the protostellar mass ($M_*$), which defines the location along an evolutionary track from a given initial condition. The properties of protostellar cores, including the protostar, disk, infall envelope, outflow, and their evolution, are calculated self-consistently from the given initial conditions. In addition to the three main physical parameters, there are also two secondary parameters: the inclination angle of the outflow to the line of sight ($\theta_\mathrm{view}$) and the level of foreground extinction ($A_V$). Thus, there are five parameters ($M_c, \Sigma_\mathrm{cl}, M_*, \theta_\mathrm{view},$ and $A_V$) that give rise to a protostellar SED. The observations are then fitted to the model grid by varying these 5 parameters and calculating the $\chi^2$ \citep[see Eq. 1 from][]{debuizer17}. Other properties, such as accretion rate, infall envelope mass, outflow cavity opening angle, and disk size, are prescribed for a given set of values of $M_c, \Sigma_\mathrm{cl}$, and $M_*$.

In the current \citet{zhang18} model grid, the main parameters are sampled as follows: $M_c$ is sampled at 10, 20, 30, 40, 50, 60, 80, 100, 120, 160, 200, 240, 320, 400 and 480\,M$_\odot$ and $\Sigma_\mathrm{cl}$ is sampled at 0.10, 0.316, 1.0 and 3.16\,$\mathrm{g\,cm^{-2}}$, which makes a total of 60 evolutionary tracks. For each track, $M_*$ is sampled at 0.5, 1, 2, 4, 8, 12, 16, 24, 32, 48, 64, 96, 128 and 160\,M$_\odot$. In the end, this yields a total of 432 physical models that have difference combinations of $M_c, \Sigma_\mathrm{cl}, M_*$. Then, for each physical model there are 20 viewing angles sampled uniformly at $\cos\theta_\mathrm{view}=0.975,0.925,\cdots,0.025$, i.e., equally distributed from 1 (face-on) to 0 (edge-on). Therefore, a total of $432\times 20=8640$ SEDs are considered. The SED models are also convolved with the transmission profiles of the instrument filters to simulate the fluxes detected in the observational bands of the various instruments. Finally, the visual extinction is constrained from 0 to a user-defined value $A_{V,{\rm max}}$. For our SED fit, we provided $A_{V,{\rm max}}=60$\,mag based on $A_{V}$ estimates of the S235~e2s3 clump \citep{dev21}. A distance of 1.65 kpc was adopted for the analysis. Table~\ref{tab1} summarizes the reliable fluxes used from the NIR to radio wavelengths for the SED modeling. Note that the data at $\lesssim8\,{\rm \mu m}$ are treated as upper limits for the SED fitting since short-wavelength fluxes at $\lesssim8\,{\rm \mu m}$ are affected by polycyclic aromatic hydrocarbons (PAHs) and thermal emission from very small grains, and these effects have not been included in the radiative transfer models \citep[see][]{zhang18}. 

\begin{figure}[t!]
\epsscale{1.2}
\plotone{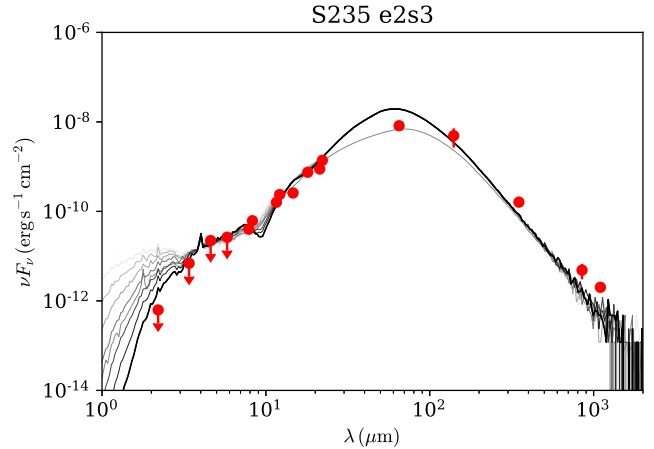}
\caption{SED of the S235\,e2s3 source using the \citet{fedriani23} fitting tool. The filled red circles are observed fluxes taken from archival data sets (see Table\,\ref{tab1}), with upper limits represented with downward-facing arrows. The gray lines represent the `good' models that satisfy the condition $\chi^2<2\chi_\mathrm{min}^2$ (gray line shades decrease with increasing $\chi^2$). The black solid line corresponds to the best-fit model. The derived best-fit model physical parameters of the source are listed in Table~\ref{tab2}.}
\label{fig5}
\end{figure}

Figure~\ref{fig5} shows results of the SED modeling with observed fluxes over wavelength. Only models that satisfy the criterion of $\chi^2<2\chi_\mathrm{min}^2$, which are considered `good' models \citep[see][for more details]{fedriani23}, were selected in the analysis. A total of 10 models satisfied this criterion and are shown as gray lines in Figure~\ref{fig5} with their shades decreasing with increasing $\chi^2$. In order to obtain the optimum values from the SED fitting, we take the average of the 10 `good' SED model parameters \citep{fedriani23}. These values then represent the best-fit model for the source, which is shown as a solid black line in Figure~\ref{fig5}. 

The computed values from the best-fit model suggest that S235~e2s3 is a protostar with a stellar mass of $6.8\pm1.2\,\Msun$ and with a total bolometric luminosity ($L_\mathrm{bol}$) of $9.63\pm2.1\times10^3\,\Lsun$. The isotropic bolometric luminosity ($L_\mathrm{bol, iso}$) i.e. without correction for the inclination and extinction is $1.68\pm0.74\times10^3\,\Lsun$. Other parameters estimated indicate the star has a disk accretion rate of 3.6$\times$10$^{-4}$\,M$_\odot\,\mathrm {yr^{-1}}$ and a foreground extinction $A_V$ of $\sim22$\,mag. Table \ref{tab2} summarizes the values of the best-fit model into two sections highlighting the SED dependent and derived parameters.

\begin{table}
\caption{SED fitting results for S235\,e2s3}              % title of Table
\label{tab2}     % is used to refer this table in the text
\centering                 % used for centering table
\begin{tabular}{ll}        % centered columns (4 columns)
\hline\hline                                    				\\[-2mm]
Parameter                             		                & 	Best-fit values$^{\mathrm{b}}$		\\
\hline  \\[-2mm]
Core mass, $M_\mathrm{c}$ ($M_{\odot}$)          			&	18.6 (17.4--19.8)  					\\ 
Mass surface density, $\Sigma_\mathrm{cl}$ ($\mathrm{g\,cm^{-2}}$)  		&	3.16 (2.16--4.16)   \\
Stellar mass, $M_*$ ($M_{\odot})$            	            &	6.8 (5.6--8.0)                    \\
Inclination angle, $\theta_\mathrm{view}$ ($^\circ{}$)  	&   78 (66--87)                          \\
Foreground extinction, $A_\mathrm{V}$ (mag)                 &	22.5 (12.8--32.2)   				 \\
\hline
%Stellar radius, $R_*$ $(R_{\odot})$                 	    &   34.7 (29.2--40.2)                   \\
Stellar age, $T_\mathrm{now}$ (yr)                  		&	3.1 (2.0--4.2)$\times10^4$          \\
Stellar temperature, $T_*$ (K)                              &   7095 (6078--8112)                   \\
Core radius, $R_\mathrm{c}$ (au)                    	    &   3670 (2500--4840)                   \\
Envelope mass, $M_\mathrm{env}$  ($M_{\odot}$) 	    	    &	5.4 (4.1--6.8)                      \\
Disk mass, $M_\mathrm{disk}$ ($M_{\odot}$)          		&	2.3 (1.5--3.2)                      \\
Disk accretion rate, $\dot{M}_\mathrm{disk}$ ($M_\odot\mathrm{\,yr^{-1}}$) 	& 3.6 (2.4--4.8)$\times10^{-4}$      \\
Bolometric luminosity, $L_\mathrm{bol}$ (\Lsun) 			&	9.63 (7.5--11.7)$\times10^3$    \\
Isotropic Bol. luminosity, $L_\mathrm{bol, iso}$ (\Lsun) 	&	1.68 (0.94--2.4)$\times10^3$    \\
Outflow opening angle, $\theta_\mathrm{w,esc}$ ($^\circ{}$) &	34 (28--40)                           \\
\hline
\end{tabular}
\begin{list}{}{}
\item[$^{\mathrm{b}}$] Mean values for $\chi^2<2\chi_\mathrm{min}^2$. The values given in the parenthesis are the ranges in the parameter values for the `good' models (see text).
\end{list}
\end{table}

\section{Discussion} \label{sec4}

NIR polarimetric observations have revealed a bipolar nebulosity around the S235\,e2s3 source. The polarization maps show a centrosymmetric pattern due to dust scattering. Molecular H$_{2}$ emission is found to display extended features along the nebulosity, tracing shocked regions of the jet. All this evidence illustrates that the S235\,e2s3 source is dynamic and presents a disk/outflow system. In this section, we discuss the outflow morphology, polarization characteristics, orientation of the outflow with respect to the local magnetic field, and the properties of the molecular H$_{2}$ emission.

\subsection{Outflow morphology and polarization characteristics} \label{polprop}

The most important results seen in our observations is the discovery of the bipolar nebulosity of an outflow (see Figure~\ref{fig2} and \ref{fig3}). The presence of bipolar outflow structures in the NIR continuum around young stars is very rare since normally they are very extincted or most of the dust in the outflow is destroyed due to the energetics of the protostar jets. Conversely, bipolar structures have been observed in several cases around post-AGB stars due to their slow mass loss leading to dusty nebulosities \citep{weintraub00, serrano20}. The visibility of structures in our NIR total intensity image could be due to longer integration time which is limited in other surveys such as 2MASS \citep{skrutskie06} and UKIDSS \citep{lawrence2007}.

The total projected length of the bipolar nebulosity spans a size of about 100,000\,au ($\sim$0.5\,pc), indicating a large extent in its dust distribution. The nebulosity in individual lobes has an extension of about 33,000\,au. Several studies of outflows from embedded Class 0/I high-mass YSOs using molecular observations have shown that they have lengths greater than 1\,pc \citep{navarete15, bally16}. However, in our case the NIR continuum observations reveal the dust in the outflow and it is not the best tracer to constrain the full length of the outflow.

The general morphology of the outflow lobes is symmetric extending in the NW-SE direction. The symmetry of the outflow and its bipolar nature has few implications. Firstly, it is reasonable to assume that the environment around the protostar is less turbulent, thus allowing for the outflow to propagate unperturbed and form a high degree of symmetry. This also indicates that the outflow is likely driven by an isolated protostar and is not located around clustered star formation. Secondly, a quiet low turbulent environment might imply particularly smooth and undisturbed accretion from the core onto the protostellar disk and the star, thus favoring a very regular and uniform occurrence of accretion. Although these assumptions are simplified, it presents an overview of the conditions around the outflow.

The outflow morphology is also useful to infer the inclination of the star system with respect to the line of sight of the observer. Since the outflow is bipolar, with the lobes being similar in size and seen easily separated spatially in the sky, we can presume that the outflow is nearly perpendicular to the observer. The derived inclination angle from the SED fitting in Section~\ref{SED} corroborates this view. The polarization levels in our observations reach a very high degree of about $70-80\%$ in the outflow nebulosity. Only few observations (for e.g. from Hubble Space Telescope) have reported similarly high values \citep{sahai98, meakin05}. Such high levels of polarization can only be measured if the scattered light from the dust particles are nearly at right angles to the line of sight of the observer. Hence, the scattering region should be perpendicular to the line of sight, therefore indicating the bipolar outflow lies approximately on the plane of sky.

\begin{figure*}[t!]
\epsscale{1.15}
\plotone{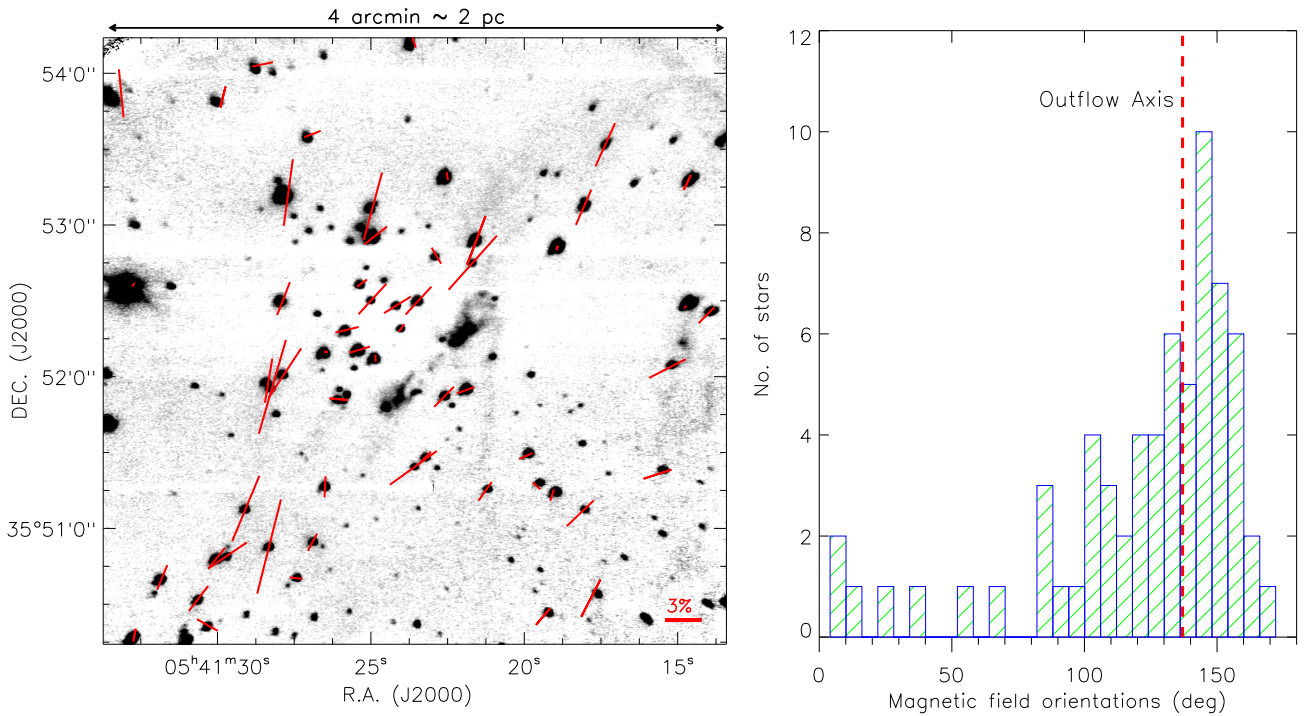}
\caption{Magnetic field map toward the S235\,e2s3 outflow region obtained from NIR background starlight polarimetry \citep{dev21}. The polarization position angles are shown as red vectors and they trace the magnetic field orientation in the plane of sky. The lengths of the vectors indicate the polarization percentage and a reference vector length is shown at the bottom left. The right panel shows a histogram of the polarization position angles representing the distribution of magnetic field orientations. The dotted red dashed line represents the position angle of the outflow axis.}
\label{fig6}
\end{figure*}
 
The polarization angles demonstrate centrosymmetric pattern indicative of single scattering mechanism in an optically thin medium. \citet{gledhill05} used polarimetric models of bipolar nebulosities to derive dust scattering properties. They found that objects with low optical depth ($\tau$) in the range $0.1<\tau<1$ tend to produce highly symmetric polarization similar to our observations. The high polarization levels are also indicative that the dust grains in the nebulosity are smaller ($<0.3\,\mu$m) than the scattering wavelength and could be made up of graphites or silicates \citep{shure95,lucas98}. At the region toward the central source, we see a small area with polarization angles perpendicular to the outflow axis. The optical depth toward the center is generally higher as these regions are extincted in the observations. The circumstellar envelope of the accreting protostar could be dominant in the central region, since a dense dust core has been identified at this position at millimeter wavelengths \citep{dewangan16, dev21}. According to the dust scattering models of \citet{whitney93, fischer94}, a polarization disk can be expected for scattered light emerging from an optically thick disk around the illuminating source. This particular case of disk polarization occurs since light passing through the disk will be scattered at least twice (above and below the plane) and loses the simple centrosymmetric polarization pattern produced by single scattering. Hence, it is likely that the disk of S235\,e2s3 is edge on to the observer, producing the observed polarization angles at the center. Again, this is consistent with the outflow being in the plane of the sky. 

It is expected that the driving source of a bipolar outflow is likely at the center of the two lobes. However, this identification may not be straightforward when several infrared sources are visible in the field which can contribute to the observed outflow structures. Hence, the identification and position of the driving source needs to be confirmed beyond what may be inferred based on the simple morphological properties and spatial distribution. We carried out polarization centroiding \citep{weintraub00} to identify the location of the illuminating source. This technique uses the scattered light to trace back the source responsible for scattering. The centrosymmetric polarization vectors in our images were rotated by $90^{\circ}$ and traced to identify their point of intersection, where the expected illuminating star is located. All the polarization vectors in our observations intersect within $3\arcsec$ of the outflow central region. This coincides with the exact position of the S235\,e2s3 source as revealed from MIR observations. Hence, it is clear that our protostar is illuminating and driving the bipolar outflow. 

\subsection{Outflow orientation with respect to local magnetic field }

Theoretical developments in star formation studies suggest that the collapse of a cloud may proceed mainly along the magnetic field orientation, since the partly ionized gas can flow easily along the field lines \citep{mouschovias1976, shu87}. At smaller scales when the collapse of the core is predominantly along the magnetic field lines, it is expected that the magnetic braking of the rotating core is more efficient when the spin axis of the core is parallel to the magnetic field \citep{mouschovias1980}. This scenario would result in the formation of a disk perpendicular to the field lines. Hence, one expects the field lines to be drawn into an hourglass shape and the outflows oriented parallel to the magnetic field \citep{galli1993}.

A number of observations toward young stars have found outflows parallel to the ambient magnetic field. These include the outflow from the Mon R2 core \citep{bally1983}, HH\,80-81 jet \citep{carr10}, IRAS 18089-1732 high-mass system \citep{beuther10}, and few low-mass Class~0 protostars \citep{chapman13}. However, more recent statistical studies of outflows using millimeter observations do not provide a consistent picture. \citet{hull13} studied the alignment of 30 cores ($\sim1000$\,au) using dust polarization maps from the TADPOL survey \citep{hull14} and found that the magnetic field lines are not tightly aligned with the outflow axis. Rather, they find the data to be consistent with scenarios where outflows and magnetic fields are if anything preferentially perpendicular. 

In Figure~\ref{fig6} we present a local magnetic field map toward the S235\,e2s3 region to compare their relative orientations with respect to the bipolar outflow. The magnetic field directions are obtained from NIR background starlight polarimetry based on dichroic extinction from dust grains aligned to the magnetic field. The complete polarimetric analysis and selection of bona fide stars for tracing magnetic field directions is presented in \citet{dev21}. We used only the sample of stars that were associated within the 2\,pc FOV of the outflow region. The results in the left panel of Figure~\ref{fig6} show that the global magnetic field orientations in the region are uniform. The morphology of the outflow also appears to be along the magnetic field lines. To compare the observations quantitatively we have shown a histogram plot of the polarization position angles in the right panel of Figure~\ref{fig6}. The histogram distribution peaks at values of about $145-150^{\circ}$. The position angle of the outflow axis determined earlier ($137^{\circ}$) is shown as a red dashed line. Both the outflow axis and the magnetic field lines have orientations within $10^{\circ}$ and indicate that the outflow is parallel to the ambient magnetic field. This result in isolation supports the general theory of core collapse along the field lines.    

\citet{jones03} found that there is a rough correlation between the outflow axis and the magnetic field orientation at the farther edge of the outflow nebulosity. They determined that when the outflow lies in the plane of sky, the magnetic field tends to be parallel to the outflow axis. This particularly matches our observations as our outflow is in the plane of sky and the field lines are traced in the surrounding medium. \citet{matsumoto06} studied the alignment of outflows with magnetic fields in cloud cores through numerical simulations. They found that the outflow tends to be aligned with the large-scale ($>5000$\,au ) magnetic field if the magnetic field in the core is larger than $80\,\mu G$. \citet{dev21} studied the magnetic field strength in the core of S235\,e2s3 and found it to be subcritical with a value of $74\pm14\,\mu G$. This value including the uncertainties is close to the threshold value of $80\,\mu G$ and may well indicate the observations match the predictions. 

One important aspect that should be considered in the context of the S235\,e2s3 outflow is the impact of the massive star in the S235 complex. Since the outflow is at the boundary of the expanding shell-like H\,{\sc ii} region, it is possible that the turbulent shock fronts and ionizing radiation from the massive star influence core collapse at its edges. In this scenario the magnetic fields are also dragged along the shell. This has been investigated by many studies \citep{kirsanova14, dewangan16, dev21} suggesting triggered star formation in this region. Hence, it is also possible that the outflow axis being parallel to the magnetic field in our case is due to the impact of the massive star. To draw a definite conclusion on such sources, it will be useful to carry out a comprehensive study of the alignment of outflows with magnetic fields in intermediate to high-mass stars and compare the results with their magnetic field strengths and outflow inclination angles. 

\subsection{\texorpdfstring{H$_{2}$}\, Luminosity and shocked emission properties}

Many studies have revealed \htwo\ outflows on scales of thousands of astronomical units to parsecs around intermediate and high-mass YSOs \citep{davis04, varricatt10, caratti15}. These observations show that the outflows are well-collimated and present similarities with outﬂows from low-mass YSOs. A key parameter in these studies is the \htwo\ luminosity and its relation to the source luminosity, which directly links the outflow properties with the driving source properties.

We studied the S235\,e2s3 outflow \htwo\ luminosity and compared it with the source properties. The \htwo\ luminosity was derived by estimating the \htwo\ 1-0 S(1) emission line flux ($F_{2.12}$) from the continuum-subtracted image. First, we calculated the total sum of pixel counts for each of the \htwo\ emission features (1a, 2a, 2b, and 2c) indicated in Figure~\ref{fig4}. A factor of 1.10 was used to multiply to the sum in order to compensate for the \htwo\ line flux that is subtracted from the {\it K} band image \citep[e.g.][]{lee19}. We get the total sum of \htwo\ emission to be 50,310 counts for an area of 272 arcsec$^{2}$. After zero-point correction and calibration, this resulted in a total 2.12$\,\mu$m line flux of $F_{2.12} = 2.28 \times 10^{-16}$ \wm\ 

Next, the 2.12$\,\mu$m \htwo\ luminosity ($L_{2.12}$) was estimated using the relation 
\begin{equation}
L_{2.12} = 4\pi{D^2} F_{2.12}\times 10^{0.4A_{2.12}}
\end{equation}
where $A_{2.12}$ is the foreground extinction at 2.12$\,\mu$m adopted as $2.3\pm1$\,mag (see Section \ref{SED}; $A_{2.12}\approx0.1 A_{V}$) and $D$ is the distance ($1.65\,\rm{kpc}$) to the source.

The total rovibrational H$_2$ luminosity ($L_{\rm H_2}$) can then be approximated from $L_{2.12}$ by assuming that the shocked emitting gas is in local thermal equilibrium (LTE). Typically, \htwo\ jets from low-mass YSOs have an average temperature of $\sim2000\,\rm{K}$, and the $L_{2.12}$ is about $10\%$ of the total $L_{\rm H_2}$ \citep{caratti06}. However, more massive jets driven by massive YSOs usually have average temperature of $2500-3000\,\rm{K}$ \citep{caratti15}, higher than low-mass jets. Therefore, $L_{2.12}$ is about $7\%$ of the total $L_{\rm H_2}$. This yields a total \htwo\ luminosity of the S235\,e2s3 outflow to be $L_{\rm H_2}=2.3_{-1.3}^{+3.5}\,\Lsun$.

This estimation has many uncertainties as we are using a general value of the extinction toward individual H$_2$ emission features, and assuming an approximate ratio of the total H$_2$ to H$_{2.12}$ luminosity. Nevertheless the estimated \htwo\ luminosity places reasonable constraints on the properties of the jet.  

\begin{figure}[t!]
\centering
\epsscale{1.18}
\plotone{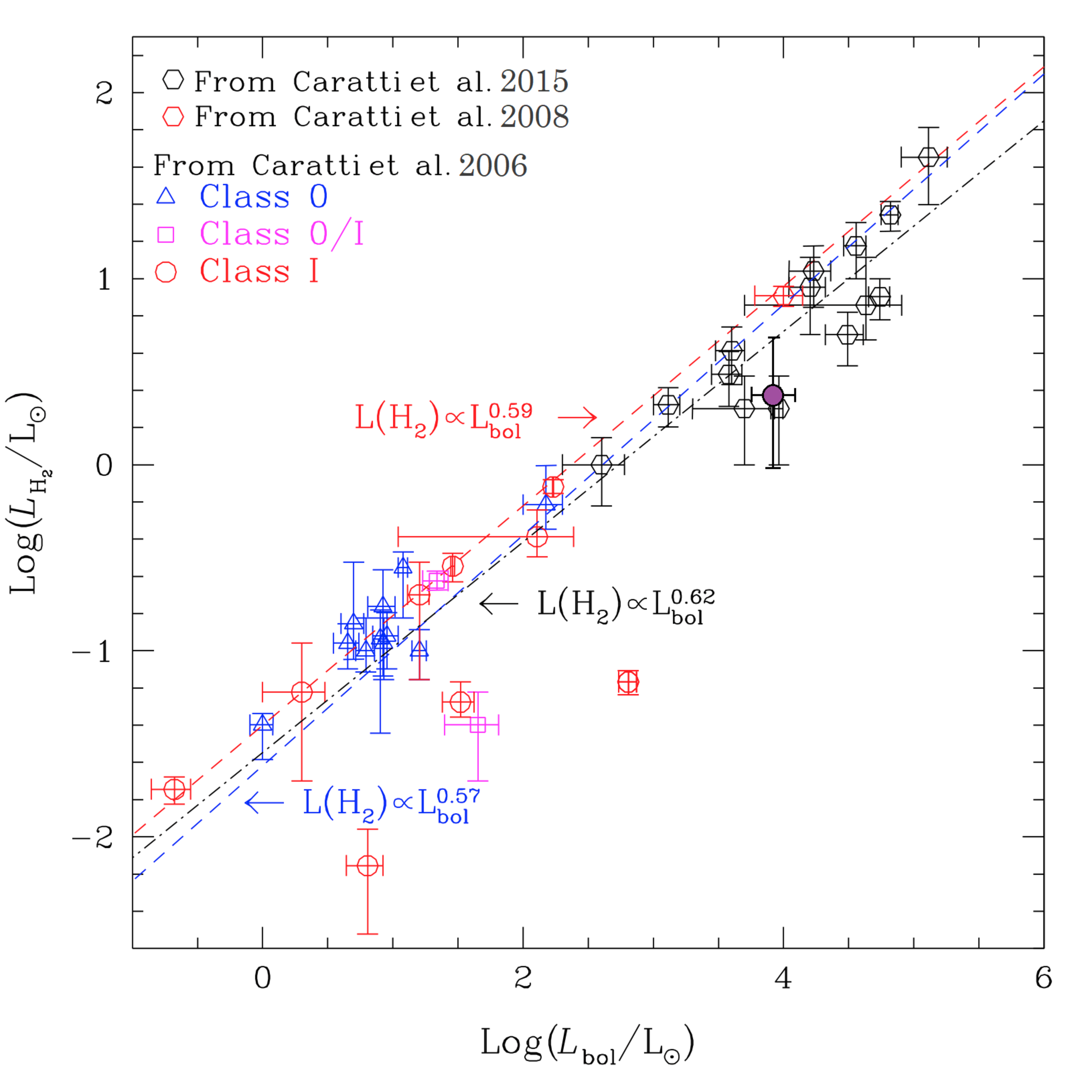}
\caption{Log ($L_{\rm H_2}$) vs. Log ($L_{\rm bol}$) from \citet{caratti15}, with results combined from previous works~\citep[][]{caratti06,caratti08}, as labeled in the upper left corner. The red dashed line indicates the fit for data points from low-mass YSOs \citep{caratti06}. The blue dashed line shows the best linear fit resulting from the sample of high-mass YSOs. The black dashed-dotted line shows the best linear fit of high-mass YSOs, excluding the five high-mass outliers \citep[see][for more details]{caratti15}. The S235\,e2s3 source is represented by a filled purple circle.}
\label{fig7}
\end{figure} 

The \htwo\ luminosities of outflows driven by low-mass YSOs are typically lower than the luminosity of the S235\,e2s3 outflow. For example, the $L_{\rm H_2}$ in \citet{caratti06} study of 23 protostellar jets driven by low-mass YSOs ranges from 0.007 to 0.76\,\Lsun. Similarly, the outﬂows detected in Serpens/Aquila from the UWISH2 survey \citep{ioan12} show observed $L_{\rm H_2}$ ranging from 0.01 to 1.0\,\Lsun. These studies are indicative that the luminosity of the S235\,e2s3 outflow corresponds to that of a higher-mass star.

Several studies show that a tight correlation exists between outflow energetics and YSO properties. \citet{caratti06} derived an empirical relationship between $L_{\rm H_2}$ and $L_\mathrm{bol}$ for a large sample of Class 0 and Class I low-mass YSOs. They found a power-law correlation of $L_{\rm H_2} \propto L^{0.58}_\mathrm{bol}$. \cite{caratti08} extended this study to include a high-mass YSO: IRAS 20126+4104 and found that the same correlation exists for this source. This relationship was then investigated for a sample of 18 massive jets from intermediate and high-mass YSOs by \cite{caratti15}. They inferred that the $L_{\rm H_2}$ and $L_\mathrm{bol}$ relationship behaves similarly as the low-mass sample and the correlation was found to be $L_{\rm H_2} \propto L^{0.62}_\mathrm{bol}$. 

\begin{figure}[t!]
\centering
\epsscale{1.18}
\plotone{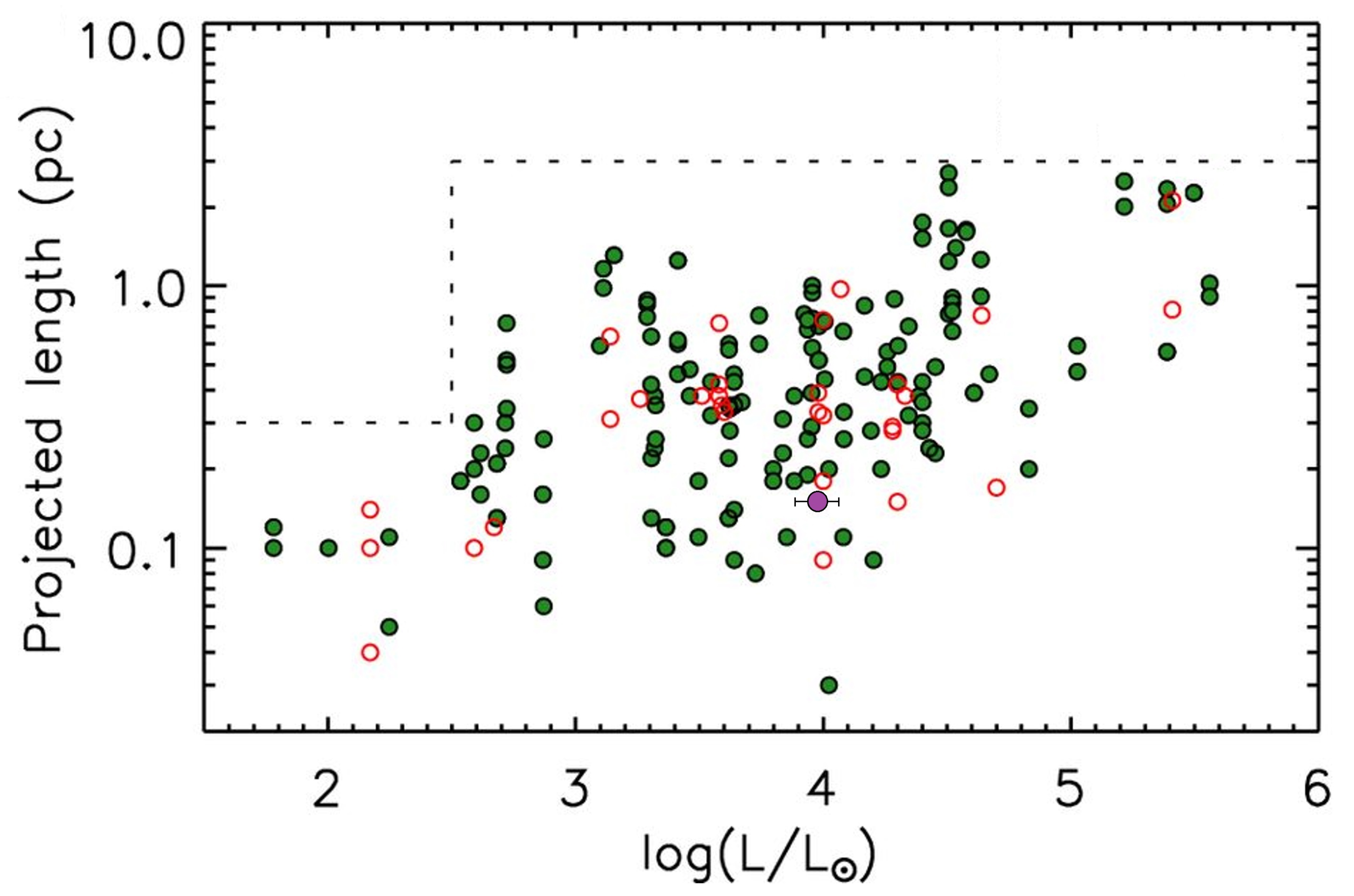}
\caption{Plot of the length of the bipolar jet as a function of the bolometric luminosity of its associated source. Values from \citet{navarete15} are shown as filled green circles while those from \citet{varricatt10} are indicated by opened red circles. Dashed lines mark the boundary of Log ($L/\Lsun$=2.5), at which there is a change in the upper limit of $l_{\rm proj}$. The values of the S235\,e2s3 source are represented by the filled purple circle.}
\label{fig8}
\end{figure}

In Figure~\ref{fig7} we show the plot of $L_{\rm H_2}$ versus $L_\mathrm{bol}$ from all the aforementioned studies. The obtained best linear fits are indicated by different colored dashed lines. The $L_{\rm H_2}$ and $L_\mathrm{bol}$ of S235~e2s3 are also shown by filled purple circle. We find that the values of S235~e2s3 is distributed close to other high-mass YSOs, matching the $L_{\rm H_2}$ versus $L_\mathrm{bol}$ relationship. An overarching trend seen from these observations is that the \htwo\ luminosity increases with the bolometric luminosity of the source, which at the protostellar stage, accounts for most or all of the source's bolometric luminosity. This suggests that the mass loss is mainly driven by the accretion power, which grows as the mass of the central source increases. In addition, it is also evident that the $L_{\rm H_2}$ versus $L_\mathrm{bol}$ is uniform across both high and low-mass protostars. Therefore, this strongly suggests that the outflows from massive YSOs are similar in nature to those from low-mass YSOs (i.e., produced by disk-mediated accretion).

The comparison between the morphology of the outflow and the intrinsic characteristics of the driving source is another important aspect to consider in the evolution of protostars. The projected length ($l_{\rm proj}$) and collimation factor ($R$ = length/width) are useful parameters in this context. In Section~\ref{h2sec} we find that the S235\,e2s3 jet has a $l_{\rm proj}$ of 31,000\,au (i.e. the distance from the source center to the farthest emission of the jet) and a width of 9900\,au. The $l_{\rm proj}$ value is generally a lower limit for the real length of the structure as the inclination angle $\theta_\mathrm{view}$ of the system affects the jet's appearance on the plane of sky. The real length of the jet can be estimated as $l_{\rm real} = l_{\rm proj}/\sin(\theta_\mathrm{view})$ which leads to $l_{\rm real} = 31700$\,au for our outflow. With these values known, the collimation factor of the S235\,e2s3 jet was estimated to be $R=3.2$.

Studies of high-velocity outflows from massive YSOs have shown collimation factors of $R=2.05$ as compared to $R=2.8$ for low-mass stars \citep{wu04}, indicating a weak tendency that outflows associated with massive stars are less collimated than those from low-mass stars \citep{richer00}. However, S235\,e2s3 has a large collimation factor suggesting that massive YSOs can also drive highly collimated outflows.

\citet{navarete15} carried out an \htwo\ survey on a large sample of massive YSOs and compared the distribution of jet projected lengths as a function of source bolometric luminosity. They also included sources from the studies of \citet{varricatt10}. Figure~\ref{fig8} shows a plot of $l_{\rm proj}$ versus $L_\mathrm{bol}$ from both these studies. We have also included the values of the S235\,e2s3 outflow represented by the filled purple circle. The length of the S235\,e2s3 jet is seen to be lower than the average length of sources with similar $L_\mathrm{bol}$. This indicates that jet lengths can vary enormously for a given stellar luminosity. The overall distribution of data points in the plot shows a large scatter, although we also find a weak trend that the largest outflows are associated with high-luminosity sources. This may suggest that generally luminous protostars drive longer jets, but it is not the main ingredient in determining the outflow length. Instead, the outflow lengths can be dependent on other factors such as the evolutionary stage of the driving source. A comparison of $l_{\rm proj}$ versus $L_\mathrm{bol}$ for different classes of YSOs will be valuable in this context. 

\section{Conclusions} \label{sec5}

We have presented a highly collimated bipolar outflow discovered from the S235\,e2s3 protostar using NIR polarimetric observations. The outflow reveals a rare bipolar dusty nebulosity in the NIR continuum images. Narrowband \htwo\ observations show shocked emission along the outflow tracing the bipolar jet's interaction with the ambient material. Based on the observations and using archival data, we have studied the properties of the outflow and derived the physical parameters of the protostar. The main results of the study are summarized as follows.

1. The bipolar outflow presents a high degree of polarization ($\sim$80\%) and reveals a centrosymmetric pattern in the polarization position angles. This suggests that the outflow nebulosity is illuminated by a single source at its center whose light is seen singly scattered at near right angles to our line of sight. Based on this, the outflow was inferred to be perpendicular to the observer. These characteristics also indicate that the S235\,e2s3 protostar is the source responsible for driving the outflow. 

2. SED fitting of the source using models of turbulent core accretion theory suggest that the S235~e2s3 protostar has a mass of $6.8\pm1.2\,\Msun$ with a total bolometric luminosity ($L_\mathrm{bol}$) of $9.63\pm2.1\times10^3\,\Lsun$. The protostar has a disk accretion rate of 3.6$\times$10$^{-4}$\,M$_\odot\mathrm{\,yr^{-1}}$ and an age of approximately $3.1\times10^4$ yr. All these parameters are consistent with the values obtained for other massive protostars.

3. The orientation of the outflow axis was found to be parallel to the ambient magnetic field direction, conforming to the expectations from magnetically induced core collapse theory. However, since the outflow is at the boundary of an H\,{\sc ii} region, feedback from the massive star might also influence the orientation of the outflow.

4. Narrowband H$_2$ ($2.12\,\mu m$) observations show multiple molecular emission features along the bipolar jet. These were identified as regions of \htwo\ flows, shocks, or knots. We estimated the total \htwo\ luminosity of the jet to be $L_{\rm H_2}=2.3_{-1.3}^{+3.5}\,\Lsun$. This value, along with the $L_{\rm bol}$, closely matched the relationship between \htwo\ luminosity and bolometric luminosity obtained for low and high-mass YSOs \citep{caratti15}. These results support the theory that jets from high-mass YSOs are similarly launched to those from low-mass YSOs via disk accretion.

5. The total length of the outflow spans about 0.5\,pc, with its collimation factor being $R=3.2$. This indicates massive YSOs can also drive highly collimated flows. Comparison of the jet length with its bolometric luminosity does not indicate any relationship between those parameters. Hence, the size of the jet generally does not scale with the mass of the protostar.

We suggest that follow-up spectroscopic observations of the protostar along with longer wavelength observations of the outflow will be very useful to understand better the dynamics and properties of this massive star disk/outflow system.

\acknowledgements{This research has been supported by the European Research Council advanced grant H2020-ER-2016-ADG-743029 under the European Union’s Horizon 2020 Research and Innovation program. D.R. acknowledges INAOE and CONACyT-Mexico for SNI grant (CVU 555629). A.L. acknowledges financial support by CONACyT-Mexico through project CB-A1S54450. A.C.G. has been supported by PRIN-INAF-MAIN-STREAM 2017 “Protoplanetary disks seen through the eyes of new-generation instruments” and by PRIN-INAF 2019 “Spectroscopically tracing the disk dispersal evolution (STRADE)”. R.F. has received funding from the European Union's Horizon 2020 research and innovation program under the Marie Skłodowska-Curie grant agreement No 101032092. We thank all the OAGH staff for their help in conducting observations with POLICAN. The authors acknowledge the use of SAOImage DS9 software which is developed with the funding from the Chandra X-ray Science Center, the High Energy Astrophysics Science Archive Center and JWST Mission office at Space Telescope Science Institute.}

\end{document}